\documentclass[12pt]{iopart}
\usepackage[utf8]{inputenc}
\usepackage{graphicx}
\usepackage{color}

\begin{document}
\title{Ab-initio analysis of superstructures revealed by STM on bilayer graphene}
\author{E. Cisternas$^1$ and J. D. Correa$^2$}
\address{$^1$Departamento de Ciencias F\'{i}sicas, Universidad de La Frontera, Casilla 54 D, Temuco, Chile}
\address{$^2$Departamento  de Ciencias F\'{i}sicas, Universidad Andres Bello, Av. Rep\'{u}blica 220, 837-0134 Santiago, Chile}
\eads{\mailto{ecisternas@ufro.cl},\mailto{jcorrea@unab.cl}}

\begin{abstract}
In this work we performed density functional theory calculations for a twisted bilayer graphene (BLG). Several
conmensurable rotation angles were analyzed and for each one a constant height mode STM image was obtained.
These STM images, calculated under the Tersoff-Hamman theory, reproduce the main features experimentally observed,
paticularly superstructures and giant corrugations.
In this way we confirm that STM characterization of twisted BLG 
can produce superstructures whose tunneling current intensity maxima occur over regions with $AA$ stacking.
Additionally we give new evidence in favour of an electronic origin for the superstructures instead
another physical grounds.
\end{abstract}

\maketitle

\section{Introduction}
\label{sec1}
The increasing interest for nanoscience and nanotechnology has evidenced the importance of the
scanning tunneling microscope (STM) to explore the nature at atomic scale. In this context graphene, 
a novel layered material with potential technological applications 
\cite{GN07}, has been intensively characterized by STM techniques and has revealed its surprising behavior:
edge states at the border \cite{KFE06,NMK06}, superstructures \cite{KCS90,LNT09,MKR10} and Van Hove Singularities 
\cite{LLL10}. The graphene particularities were implied before by STM characterization of the 
(0001) surface of highly oriented pyrolitic graphite (HOPG): difficulties to obtain STM complete
atomic resolution as well the observation of superstructures are examples of this.

Extending the preceding statements, let us precise that graphene is a stable bidimensional
structure formed by carbon atoms ordered in a hexagonal lattice (the first neighbor distance is $1.42$ \AA). 
In crystalline graphite (HOPG) such layers are stacked on each other in a $AB$ sequence, 
and therefore it is possible to identify two type of atomic sites: the $\alpha$-type sites, corresponding to atoms 
with neighbors directly above and below in adjacent layers; and the $\beta$-type sites, 
which do not have direct neighbors in the adjacent layers. An important consequence of the crystalline structure of 
HOPG is that near the Fermi level the bulk presents a Local Density of States (LDOS) which is larger
for atoms on $\beta$-type sites \cite{TL88}. The theory of STM image formation \cite{tersoff-hamann,Selloni}
considers that the accessible states by STM are those in the energy window $[E_{Fermi}-e V_{bias},E_{Fermi}]$.
Consequently in HOPG only those atoms on $\beta$-type sites appears 
visible for a STM operating at low bias voltages and they form the triangular lattice usually reported on experiments. However the LDOS of $\alpha$
and $\beta$ sites tends to be equal by increasing the bias voltages and the persistence of the triangular lattice
at higher bias voltages has been explained by resolution losses introduced for the STM tip size \cite{CSF09}.

\begin{figure}[h]
\includegraphics[width=.8\linewidth]{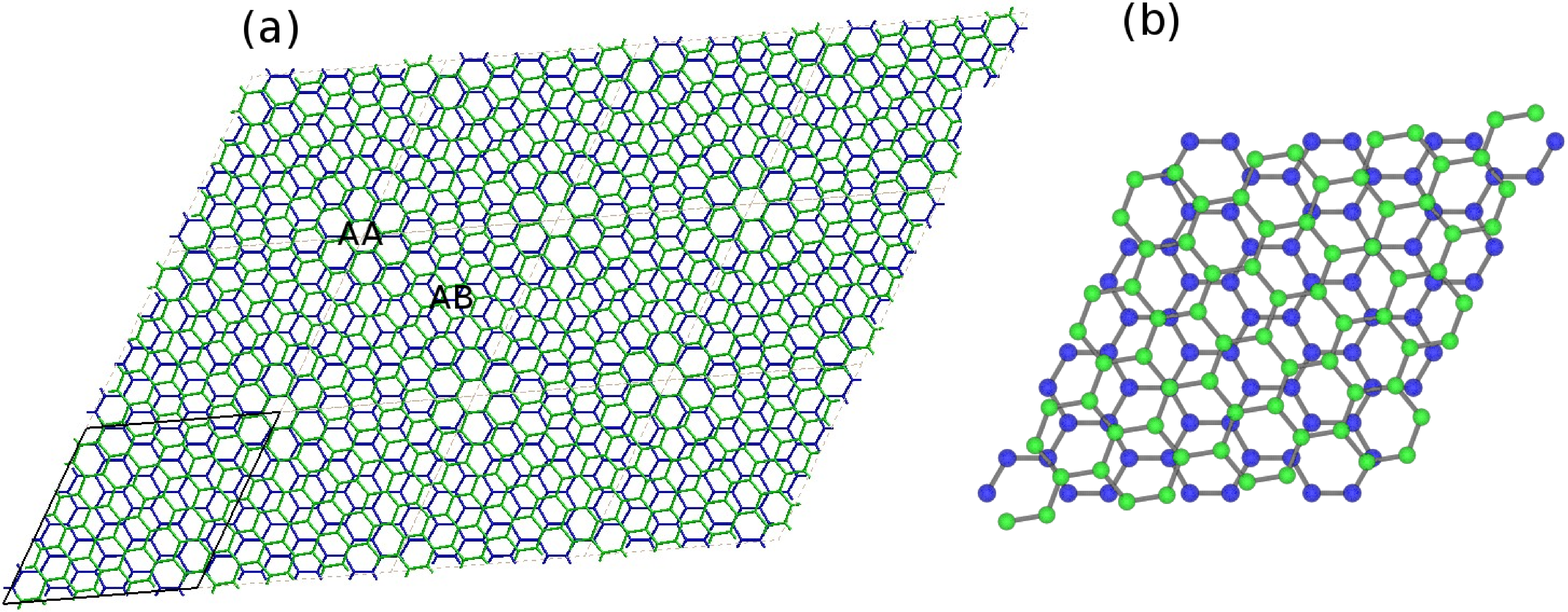}
\caption{(a) Scheme view of twisted bilayer graphene for $7.3^\circ$ as rotation angle. 
(b) Magnification of the unit supercell which expands the superstructure. 
Green sticks are for top layer and the blue ones for bottom layer.
\label{figure0}}
\end{figure}

Following the preceding ideas, the observation of superstructures during STM experiments can be explained by considering a relative rotation among the constituent graphene 
layers. This is because rotations generate \emph{Moir\'e Patterns} where it is possible
to identify regions with a high concentration of $\beta$-type sites, called $g-\beta$ sites,
and regions with a high concentration of $\alpha$-type sites, called $g-\alpha$ or $AA$ stacked regions [see Figure \ref{figure0}(a)].
In both cases $g$ stands for \emph{giant} \cite{XSG93}.
In this frame the $g-\beta$ regions would appear brilliant in STM images due to the large LDOS of atoms on $\beta$ sites \cite{XSG93}.
However, Rong and Kuiper \cite{RK93}, on base of first principles DOS calculations \cite{Charlier}, 
proposed that brilliant zones occur over the $g-\alpha$ regions.
Aditionally a third explanation for the superstructures considers that more important than the electronic effects induced by rotation
between weakly interactive graphene layers, could be surface deformations
introduced by STM tip while it scans the surface (see Ref. \cite{PD05} and references therein).
This has been a long standing controversy which has attracted renewed interest 
over the STM image formation of graphite surface \cite{CSF09,ZP08,KC09,WDC09} and over the physical origin 
of superstructures \cite{PD05,CSS07,CFV08,TMM10}.

In this context, an important tool to analyze the effect of rotation among graphene layers is the calculation of
STM images and the corresponding contrast with experimental data. Following this objective
we have focused in the task of generate STM images for different rotation angles between two graphene layers.
We have organized this paper as follows: details of the images calculation are given in next section;
theoretical results are summarized in section \ref{sec3} and finally the main conclusions are presented.

\section{Calculation Method}
\label{sec2}

The system under study corresponds to BLG presenting a relative rotation angle between its layers 
(twisted bilayer graphene). Such rotation occurs around the stacking direction and despite for any angle one can 
identify a superstructure (or moir\'e pattern) only for very particular angles the misoriented layers are in 
commensuration \cite{TMM10,KC05,SSK10}. A commensurable unit supercell
requieres a rotation from a vector $\vec{V}_1=m\vec{a}_1+n\vec{a}_2$ to $\vec{V}_2=n\vec{a}_1+m\vec{a}_2$, where 
$\vec{a}_1=(\sqrt{3},-1)a_0/2$ and $\vec{a}_2=(\sqrt{3},1)a_0/2$ are the graphene basis vectors; $m$ and $n$ are 
integers and $a_0=2.46$ \AA \, is the lattice constant.
The commensurable rotation angle is defined by

\begin{equation}
\cos{\theta}=\frac{m^2+4mn+n^2}{2(m^2+mn+n^2)} \, ,
\end{equation}

\noindent and the unit super cell vectors correspond to: $\vec{t}_1=\vec{V}_2=n\vec{a}_1+m\vec{a}_2$ and $\vec{t}_2=-\vec{a}_1+(m+n)\vec{a}_2$ \cite{TMM10}.
The unit supercell contains $N=4(m^2+mn+n^2)$ atoms and its periodicity results $D=a_0/[2 \sin(\theta/2)]$ 
\cite{XSG93,PD05}. Thus, commensurable unit supercells are complety defined by the index $m$ and $n$ and those
selected to this study appear in Table \ref{table1} with their corresponding number of atoms $N$, rotation angle 
between layers $\theta$ and periodicity $D$. For example, the unit supercell for twisted bilayer graphene with the angle $7.3^\circ$
is shown in Figure \ref{figure0}(b).

\begin{table*}[h]\begin{center}
\caption{Selected commensurable unit supercells and their corresponding number of
constituent atoms ($N$), rotation angle ($\theta$) and periodicity ($D$). The last column shows periodicities experimentally reported.}
\begin{indented}
\begin{tabular}{ccccccc}
\br
	m&n&N&$\theta$& D (nm) &\multicolumn{2}{c}{D Exp. (nm)}\\
\hline \hline
    2&1&28 &$21.8^\circ$&$0.65$& $0.65$ Ref. \cite{LLR11}&\\ 
    3&2&76 &$13.2^\circ$&$1.05$& $1.06$ Ref. \cite{NRF93} &$0.95$ Ref. \cite{MPF08}\\ 
    4&3&148& $9.4^\circ$&  $1.50$ & $1.50$ Ref. \cite{EW88}& $1.50$ Ref. \cite{SKB02}\\ 
    5&4&244& $7.3^\circ$&  $1.91$ & $1.76$ Ref. \cite{AMN88}&$1.71$ Ref. \cite{MAY98} \\ 
    6&5&364& $6.0^\circ$&  $2.30$ & $2.20$ Ref. \cite{THI00}&$2.40$ Ref. \cite{YBG92}  \\ 
\br
\label{table1}
\end{tabular}
\end{indented}
\end{center}
\end{table*}

Our DFT calculations were carried out using the SIESTA \textit{ab initio} package 
\cite{Siesta} which employ norm-conserving pseudopotentials and localized atomic 
orbitals as basis set. Double-$\zeta$ plus polarization functions were used under the local  density 
approximation \cite{LDA}. All structures were fully relaxed until the atomic forces are smaller than 0.02~eV/\AA. 
We consider supercells with periodic boundary conditions for the plane layers, while  
in the perpendicular direction to the layers plane the separation between slabs is fixed to 14~\AA. 
The Brillouin zone sampling was performed using a Monkhorst-Pack mesh of $10\times10\times1$. 

The images were obtained using the code \emph{STM} 1.0.1 (included in the SIESTA package).
This code uses the wavefunctions generated by SIESTA and computed on a reference
plane and extrapolates the value of these wavefunctions into vacuum.
Such reference plane must be sufficiently close to the surface so that the charge density is large and well described.
The STM data is generated under the Tersoff-Hamann theory \cite{tersoff-hamann}, 
while data visulization was possible using the WSxM 5.0 freeware \cite{HFG07}. Gaussian smooth was performed to obtain
the final STM image.

\section{Results}
\label{sec3}

\begin{figure}[h]
\includegraphics[width=.8\linewidth]{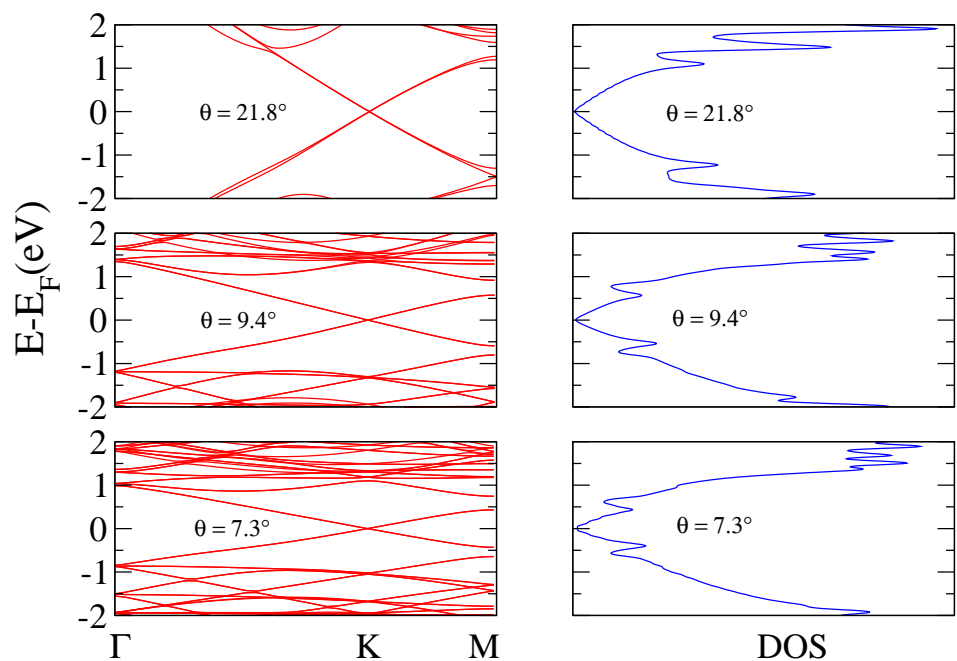}
\caption{Bands structure  and total  density of states for different twisted bilayer of graphene.
Left panels show the bands structures and right panels show  the total  density  of states.
\label{figure3}}
\end{figure}

Electronic properties and constant height mode STM images were calculated for the five commensurable 
supercells presented in Table \ref{table1}. Figure \ref{figure3} shows band structure and total density of 
states (DOS) for three of these twisted bilayer graphene. Clearly, a decrease of the rotation angle $\theta$ diminishes the 
energy difference between the VHS near the Fermi Level. This effect was reported in previous works 
\cite{LLL10,TMM10,SCV10} and has its origin in the renormalization of Dirac electrons velocities, which is observed in 
the band structures as a decreasing in the slope of the Dirac cone. Additionally the VHSs show the well known 
asymmetry between electrons and holes, which is observed as a difference in the height of the VHSs. These theoretical 
predictions have been recently corroborated by STM/STS experimental measurements on twisted graphene 
layers with $\theta=1.79^\circ$ \cite{LLL10} and ensure us that our calculations reproduce the main
electronic characteristics of the twisted bilayer graphene. 

\begin{figure}[h]
\includegraphics[width=.8\linewidth]{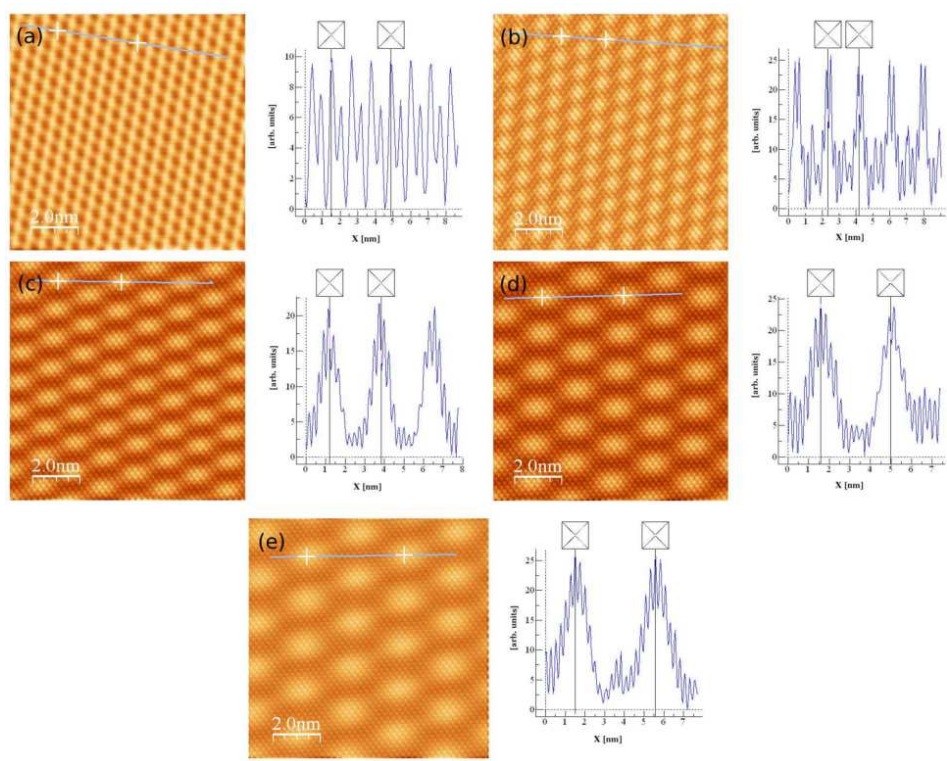}
\caption{Constant height mode STM images calculated for the five commensurable unit supercells
presented in Table \ref{table1}. The line profiles (along the line indicated on the corresponding STM image)
appears at right of each one.
\label{figure1}}
\end{figure}

As the occupied states contributing to the STM image are in the energy window [$E_{Fermi}-e|V_{bias}|,E_{Fermi}$] 
\cite{tersoff-hamann,Selloni}, the STM image bias voltage dependence was studied by considering several values in the 
range  $0.05$ V $< V_{bias} < 3.0$ V. As example, Figure \ref{figure1} shows calculated STM images  and their 
corresponding line profiles for the case $V_{bias} = 1.0$ V and the tip-to-surface distance $d_{TS} = 1.0$ \AA. These 
calculated STM images show bright zones indicating tunnel current maxima, which form triangular structures with 
periodicities much larger than the graphite lattice constant. These \emph{superlattice constants} correspond to the 
periodicities $D$ shown in Table \ref{table1}. Another important result derived from these calculations is that the 
brilliant zones are over the $AA$ stacked regions in accordance with the proposal of other authors: Rong and Kuiper 
\cite{RK93} and Campanera et al. \cite{CSS07}, who based their conclusions in DOS calculations for different stacking 
sequence of graphite \cite{Charlier}; and Trambly de Laissardiere et al. \cite{TMM10}, who combined tight-binding with 
ab-initio calculations to study the LDOS on twisted bilayer graphene.

\begin{figure}[h]
\includegraphics[width=.65\linewidth]{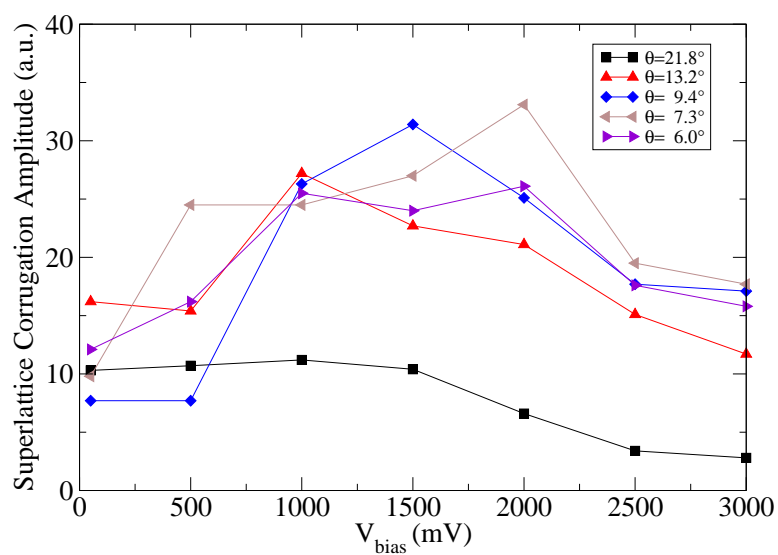}
\caption{Superlattice Corrugation Amplitude on calculated STM images as a function of bias voltages.
The rotation angles for the twisted BLG are indicated in the figure.
\label{figure4}}
\end{figure}

Line profiles were performed along the segment indicated in each STM image of Figure \ref{figure1}.
Although the calculated images correspond to constant height mode, the line profiles might be compared with those 
experimentally obtained in constant current mode assuming that depressions in the charge density lead to proportional 
approximations of the tip-to-the-surface in order to keep the tunnel current constant. So, our results show two 
important features revealed also during the STM experiments: a roughnes presenting the graphite periodicity and a 
large oscilation amplitude presenting the superlattice periodicity \cite{XSG93,RK93,PD05}.
For the first case, the roughnes reflects the electronic charge accumulation at the atomic sites, which is detected by 
STM tip. For the second case, the oscillation amplitude is much more larger than the roughness due to the atomic sites 
and it has been named  ``giant'' or ``superlattice corrugation'' \cite{PD05}. Clearly our calculations suggests that the physical origin of 
the \emph{superlattice corrugation} is mainly an electronic effect and consequently the wide gamma of alternative 
explanations (see \cite{PD05} and references therein) seems to be less important at least for STM images of BLG.
Also we found, coinciding with the experimental observations on graphite surfaces \cite{RK93,PD05}, that bias 
voltage has a strong influence in the \emph{Superlattice Corrugation Amplitude} (SCA).
Figure \ref{figure4} shows SCA obtained from calculated STM for the rotation angles under study and with 
different bias voltages. 
The SCA corresponding to $\theta=21.8^\circ$ is the smaller one and presents a current maximum at $\approx 1.0$ V. 
Such SCA maximum seems to be a consequence of the DOS which, in the case  $\theta = 21.8^\circ$, shows the first VHS at $\approx 
-1.0$ V.
Of course the main contribution to this VHS comes from the Local Density of States (LDOS) over the $AA$ stacked 
regions, but the decrease of SCA for $V_{bias} > 1.0$ V is evidence that from that value other regions start to 
contribute to the DOS. For lower rotation angles, as i.e. $9.4^\circ$ and $7.3^\circ$, 
the first VHS occurs closer to the Fermi Energy (see Fig. \ref{figure3}) and its main contribution comes from the $AA$ stacked regions.
However when the bias voltage increases over 2.0 V the LDOS from other regions start to contribute equally and SCA starts to 
diminish. Such behavior for the LDOS was also experimentally revealed by STS on the bright and dark 
regions of STM data \cite{LLL10,LLR11}.

On the other hand the STM dependence on the tip-to-sample distance ($d_{TS}$) was also studied by considering
different values in the range $1.0$ \AA \, $< d_{TS} <$ $2.0$ \AA.   
Thus, the present STM image generation methodology shows that the tip-to-surface distance has no 
consequence in the calculated STM images and current intensity maxima remain over the regions with large concentration 
of $\alpha$-sites ($AA$ regions). This result indicates that the STM image generation methodology is well defined in 
the $d_{TS}$ range considered, but does no give definitive conclusions about tip-to-sample distance dependences as 
those theoretically predicted \cite{CFV08} and experimentally observed \cite{MKR10} for trilayer graphene case.

\section{Conclusions}

From ab-initio methods we have calculated constant heigth mode STM images for twisted graphene bilayer presenting five commensurable rotation angles.
The calculated images show superstructures as those revealed during STM characterization of graphite surface and reproduce
their main features. In particular the line profiles, obtained from the calculated images,
show two important features experimentally observed: a large oscillation amplitude (giant corrugation) with the periodicity of the superstructures;
and a roughness with the graphene lattice periodicity. Thus, these results are showing that the superstructures have a strong electronic origin
and, therefore, mechanical tip-sample interaction, or another physical grounds, are not always relevant to explain the phenomena.
  
Three additional important conclusions derived from this study must be remarked: first, despite the weak interlayer interaction, a relative 
rotation between them can modify considerably the surface electronic structure; second, results show clearly that, for 
twisted bilayer graphene, the current intensity maxima appear over $AA$ stacked regions; and third, the bias 
voltage has a strong influence in the \emph{superlattice corrugation} amplitude.

\ack
This work was partially supported by Universidad de La Frontera,
under Project DI11-0012.  
J. D. acknowledges FONDECYT postdoctoral program under Grant No. 3110123 
We thanks Centro de Modelaci\'on y Computaci\'on Cient\'ifica (CMCC), Universidad de La Frontera,
by computational facilities and programing help. 
Dr. Marcos Flores is gratefully acknowledege for useful discussions.

\newpage

\end{document}